\documentclass[fleqn,twoside]{article}
\usepackage{espcrc2}
\usepackage{epsfig}
\usepackage{amssymb}
\usepackage[figuresright]{rotating}

\newcommand{\AmS}{{\protect\the\textfont2
  A\kern-.1667em\lower.5ex\hbox{M}\kern-.125emS}}

\def\gtap{\;\raisebox{-.5ex}{\rlap{$\sim$}} \raisebox{.5ex}{$>$}\;}

\title{Spectral Functions, Maximum Entropy Method and Unconventional Methods
in Lattice Field Theory}

\author{Chris Allton, Danielle Blythe and Jonathan Clowser\\
\vspace{3mm}
Department of Physics, University of Wales Swansea,
Singleton Park, Swansea SA2 8PP, United Kingdom}

\begin{document}

\begin{abstract}

We present two unconventional methods of extracting information from
hadronic 2-point functions produced by Monte Carlo simulations.  The
first is an extension of earlier work by Leinweber which combines a
QCD Sum Rule approach with lattice data. The second uses the Maximum
Entropy Method to invert the 2-point data to obtain estimates of the
spectral function.  The first approach is applied to QCD data, and the
second method is applied to the Nambu--Jona-Lasinio model in (2+1)D.
Both methods promise to augment the current approach where physical
quantities are extracted by fitting to pure exponentials.

\end{abstract}

\maketitle



\section{Introduction}

The conventional approach of studying lattice hadronic correlation
2-point functions, $G_2(t)$, is to fit the large time, $t \ge t_{min}$,
interval of $G_2(t)$ to an exponential, $Z e^{-Mt}$, where $M$ is the
mass of the asymptotic state, and $Z$ is the overlap of that state
with the interpolating operator used. This method can be extended by
including further exponentials, $Z_i e^{-M_i t}$, which correspond to
the lowest excited states of the channel in question. However,
fundamentally, the small time part of the correlation function,
$t < t_{min}$, is always thrown away. This paper presents two approaches
which are able to include the $t\approx 0$ region of $G_2(t)$ in
the analysis.

The first follows the approach of Leinweber\cite{lw} where a QCD Sum
Rule technique is used to model the short time behaviour of
$G_2(t)$. This is detailed in the next section where new results for
non-degenerate mesons are presented, and in
Section 3 where these calculations are extended to include lattice
artefacts coming from the Wilson action.

The second approach uses the ``Maximum Entropy Method'' of numerical
analysis to invert $G_2(t)$ into its spectral function representation.
Section 4 describes this method and its application to the 
Nambu--Jona-Lasinio model in (2+1)D.



\section{QCD Sum Rule approach and non-degenerate mesons}

In \cite{lw} a method was presented which combines two of the
main non-perturbative approaches to strongly coupled field
theory: the lattice and QCD Sum Rules. This section presents
an overview of this approach and its extension to the case of
non-degenerate mesons with non-zero momentum.

The basic quantity considered is the short distance expansion (OPE) of the
quark propagator, $S_q(x)$,
\begin{equation}
S^{OPE}_q(x) =
           \frac{\gamma \cdot x}{2 \pi^2 x^4} +
           \frac{m}{(2 \pi x)^2}            -
           \frac{\langle : \overline{q}q : \rangle}{2^2 3}
           + \cdots,
\label{eqn:qprop}
\end{equation}
where $m$ is the quark mass.

The 2-point correlation function at zero momentum, $G_2(t)$,
can be written as,
\begin{equation}
G_2^{\Gamma}(t) =
\sum_{\underline{x}} \langle 0| {\cal T} \{J_{\Gamma}(\underline{x},t)
                          \overline{J}_{\Gamma}(\underline{0},0)\}|0\rangle,
\label{eq:g22}
\end{equation}
or, Wick contracting, as
\begin{equation}
G_2^{\Gamma}(t) =
\int d^3x Tr \left\{S_q(x)\Gamma S_q(-x)\Gamma\right\}.
\label{eq:g2}
\end{equation}
$\Gamma$ is a spinor matrix which defines the quantum numbers of the
interpolating operator, $J_{\Gamma}$, and thereby of the channel
considered. In Eq.(\ref{eq:g2}) we have restricted our attention
to mesons (although in \cite{lw} baryons were considered).

The QCD Sum Rules approach (and the method of \cite{lw})
substitutes the short distance expansion $S_q^{OPE}$ 
into the expression for $G_2$ in Eq.(\ref{eq:g2}).
This defines $G_2^{OPE}$.

We now introduce the spectral density, $\rho(s)$, which is defined
via the Laplace transform of $G_2(t)$,
\begin{equation}
G_2(t) = \int_0^\infty \rho(s) e^{-st} ds.
\end{equation}
The spectral density, $\rho(s)^{OPE}$, corresponding to $G_2^{OPE}$, is
obviously valid for large values of $s$ (i.e. short distances).
We cannot expect $\rho(s)^{OPE}$ to reproduce the non-perturbative
behaviour of QCD which generates bound states. The QCD Sum Rules
``Continuum Model'' allows for this by introducing a ``continuum threshold'',
$s_0$, above which $\rho(s)^{OPE}$ is assumed valid. The non-perturbative,
confining features of the system are modelled with a delta function
at $s = M < s_0$. The full spectral function is,
\begin{equation}
\rho(s) = Z \delta(s-M) + \xi \; \theta(s-s_0) \rho(s)^{OPE}.
\label{eq:rho}
\end{equation}
$Z$ is the normalisation of the ground state whose mass is $M$.
and $\xi$ a parameter which takes account of the different normalisations
of the lattice versus continuum states (see \cite{lw}).

For completeness, we write here the two-point function that
Eq.(\ref{eq:rho}) defines:
\begin{equation}
G_2(t) = Z e^{-Mt} + \xi 
\int_{s_0}^\infty \rho(s)^{OPE} e^{-st} ds.
\label{eq:g2_final}
\end{equation}

The above method was first employed in the study of lattice data in
\cite{lw} where the analytic expressions for Eq.(\ref{eq:g2_final})
were derived in the case of zero-momentum baryonic channels.
In \cite{orange}, this work was extended to zero-momentum mesons and the
continuum limit of the Ansatz, Eq.(\ref{eq:g2_final}), was studied.
In \cite{abc}, we extend this approach still further by studying non-degenerate
mesons at both momentum $\vec{p} = \vec{0}$ and $\frac{1}{\sqrt{3}}(1,1,1)$.
We show here the expression
for $G_2^{OPE}$ in the case of the (spatial) vector channel
at zero momentum.
\begin{eqnarray}  \nonumber
G_2^{Vi}(t)^{OPE,cont} \!\!\!&=&\!\!\! \frac{-1}{2 \pi^2 t^3} \\
 \!\!\!&+&\!\!\! \frac{3}{8 \pi^2 t} [ m_1^2 + m_2^2 - 2 m_1 m_2 ]
\label{eq:vi_hl}
\end{eqnarray}
where $m_{1,2}$ are the two valence quark masses of the meson, and the
superscript $cont$ refers to the fact that the continuum expression
for $S_q$, Eq.(\ref{eqn:qprop}), was used in the derivation of
Eq.(\ref{eq:vi_hl}).

In this work the Monte Carlo data for $G_2(t)$ was generated using UKQCD
computing resources.  Table 1 lists the lattice
parameters used in this simulation.

\begin{figure}[htb]
\centering
\vspace{1cm}
\includegraphics[scale=0.29]{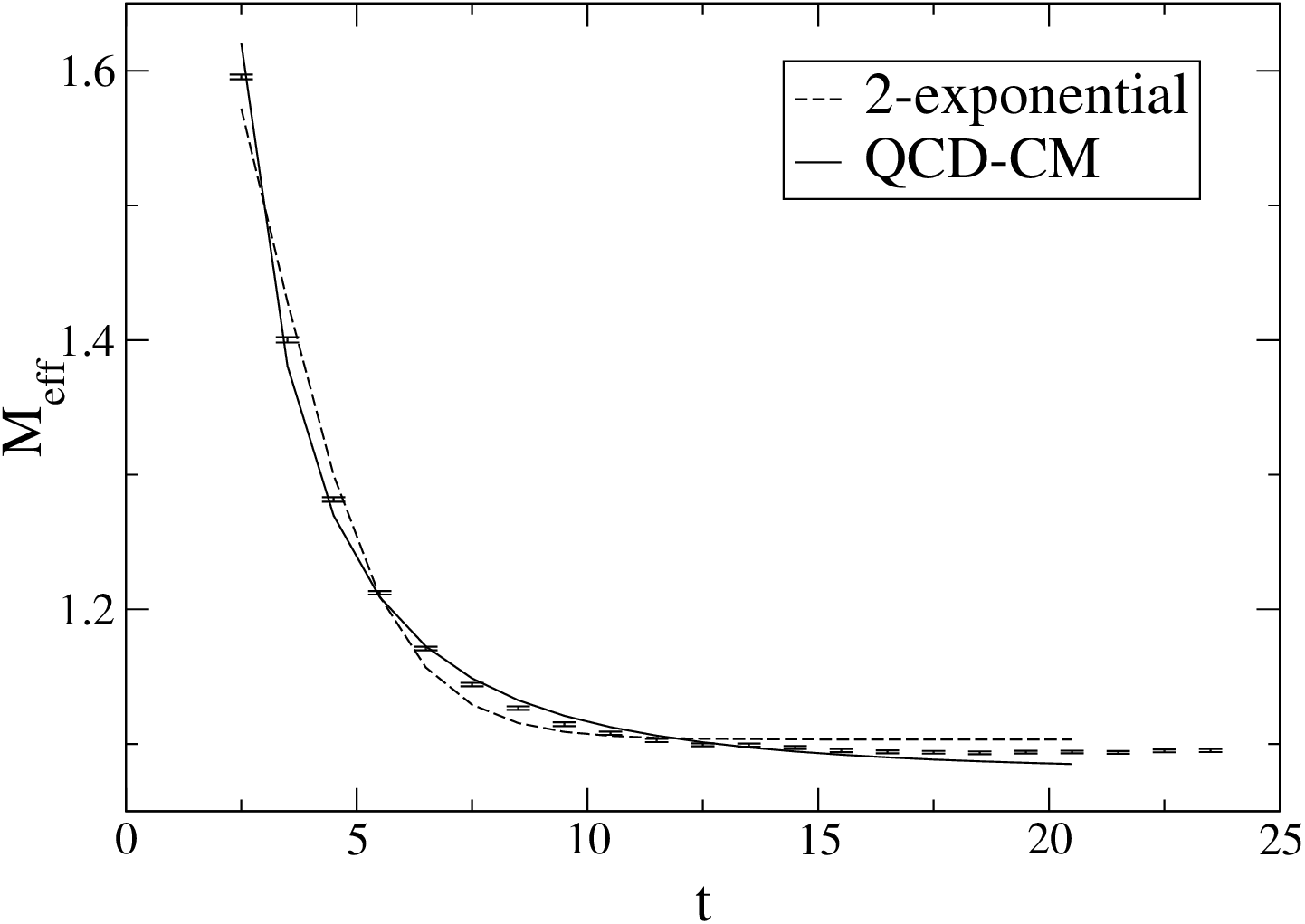}
\caption{\small 
Plot of the effective mass for the $\kappa = 0.1385$ case.
The result of a 2-exponential fit together with the QCD Continuum
Model fits of Eq.(\ref{eq:vi_hl}) \& Eq.(\ref{eq:vi_wf}) are shown.
(The latter two fits are indistinguishable on the plot by eye.)
}
\label{fig:vi_hl}
\end{figure}

In Figure \ref{fig:vi_hl} we show the effective mass for the vector
channel at $\kappa_{1,2} = 0.1385$. In this plot the QCD Continuum
Model fitting function
Eq.(\ref{eq:vi_hl}) is shown together with the result of a
2-exponential fit. The fitting range for both Eq.(\ref{eq:vi_hl}) and
the 2-exponential fit was $t=2-20$.

As can be seen from this plot, the QCD Continuum Model fit,
Eq.(\ref{eq:vi_hl}), clearly gives a better fit. This can be seen
quantitatively in Table \ref{tb:fits} where the $\chi^2$ values for
both fits are shown together with the estimates of the $Z$ and $M$
values for a wide range of $\kappa$ values. The results of a
1-exponential fit (using $t=12-20$) are shown as a control.



\begin{table}
\begin{center}
\begin{tabular}{ll}
\hline
Parameter & Value \\
\hline
$\beta$ & 6.0 \\
$N_f$   & 0 (i.e. Quenched) \\
Volume  & $16^3 \times 48$ \\
$c_{SW}$    & 0 $\;\;\;$ (i.e. pure Wilson Action) \\
\hline
\end{tabular}
\label{tb:lattice_params}
\end{center}
\caption{\small The lattice parameters used.}
\end{table}



\begin{table}[*htbp]
\begin{tabular}{llll}
\hline
 &&&\\
Fit  & Z & M & $\chi^2$/d.o.f. \\
Type &   &   &                 \\
\hline
\multicolumn{4}{c}{$\kappa = 0.1550$} \\
1-exp         & 6.7( 5)$\times 10^{ -3}$ & .426( 4) & - \\
2-exp         & 8.5( 2)$\times 10^{ -3}$ & .440( 2) & 160/15 \\
CM            & 5.9( 2)$\times 10^{ -3}$ & .419( 3) & 30/15 \\
${\cal O}(a)$ & 5.9( 2)$\times 10^{ -3}$ & .419( 3) & 30/15 \\
 &&&\\
\hline
\multicolumn{4}{c}{$\kappa = 0.1540$} \\
1-exp         & 8.6( 3)$\times 10^{ -3}$  & .467( 3) & - \\
2-exp         & 1.03( 2)$\times 10^{ -2}$ & .479( 2) & 210/15 \\
CM            & 7.6( 2)$\times 10^{ -3}$  & .461( 2) & 36/15 \\
${\cal O}(a)$ & 7.6( 2)$\times 10^{ -3}$  & .461( 2) & 36/15 \\
 &&&\\
\hline
\multicolumn{4}{c}{$\kappa = 0.1530$} \\
1-exp         & 1.05( 3)$\times 10^{ -2}$ & .509( 2)  & - \\
2-exp         & 1.25( 2)$\times 10^{ -2}$ & .5193(13) & 240/15 \\
CM            & 9.3( 2)$\times 10^{ -3}$  & .5021(15) & 40/15 \\
${\cal O}(a)$ & 9.3( 2)$\times 10^{ -3}$  & .5022(15) & 40/15 \\
 &&&\\
\hline
\multicolumn{4}{c}{$\kappa = 0.1455$} \\
1-exp         & 3.00( 4)$\times 10^{ -2}$ & .8171(10) & - \\
2-exp         & 3.44( 4)$\times 10^{ -2}$ & .8255(10) & 170/15 \\
CM            & 2.47( 5)$\times 10^{ -2}$ & .8073(11) & 44/15 \\
${\cal O}(a)$ & 2.51( 4)$\times 10^{ -2}$ & .8080(11) & 39/15 \\
 &&&\\
\hline
\multicolumn{4}{c}{$\kappa = 0.1420$} \\
1-exp         & 4.13( 5)$\times 10^{ -2}$ & .9576( 9) & - \\
2-exp         & 4.70( 5)$\times 10^{ -2}$ & .9655(10) & 136/15 \\
CM            & 3.17( 8)$\times 10^{ -2}$ & .9450(12) & 49/15 \\
${\cal O}(a)$ & 3.30( 7)$\times 10^{ -2}$ & .9466(11) & 41/15 \\
 &&&\\
\hline
\multicolumn{4}{c}{$\kappa = 0.1385$} \\
1-exp         & 5.35( 7)$\times 10^{ -2}$ & 1.0958( 9) & - \\
2-exp         & 6.04( 7)$\times 10^{ -2}$ & 1.1033(10) & 115/15 \\
CM            & 3.68(15)$\times 10^{ -2}$ & 1.0792(14) & 58/15 \\
${\cal O}(a)$ & 4.05(10)$\times 10^{ -2}$ & 1.0826(11) & 45/15 \\
 &&&\\
\hline
\end{tabular}
\caption{\small Values for $\chi^2$, $Z$ and $M$ for the
vector channels for a single exponential fit (used as a control),
a 2-exponential fit, the ``QCD Continuum Model'' (CM) fit to Eq.(\ref{eq:vi_hl}),
and the ``lattice-QCD Continuum Model'' (${\cal O}(a)$) fit using Eq.(\ref{eq:vi_wf}).
All fitting ranges were $t=2-20$ with the exception of the single
exponential fit which was $t=12-20$.}
\label{tb:fits}
\end{table}



\section{Analytic lattice results for non-degenerate mesons}

In Section 2, an analytic expression for the 2-point function $G_2$
was derived by using the continuum, short-distance expansion for the
quark propagator, $S_q$. In this section we extend this work still
further using the {\em lattice} expression for (the co-ordinate space)
quark propagator, $S_q(x)^{LAT}$, recently derived by Paladini for Wilson
quarks\cite{paladini}.

While the momentum space expression for lattice propagators has a
simple, closed form, the co-ordinate space representation does not.
Before the work of \cite{paladini}, $S_q(x)^{LAT}$
was known in two cases only: the massless limit \cite{lu_we};
and the $x=0$ limit \cite{bcp}.

In \cite{paladini}, $S_q(x)^{W}$ (for Wilson Quarks) was derived as a
continuum expansion (i.e. as a Taylor expansion in the lattice
spacing, $a$). To obtain this expression the authors of \cite{paladini}
used a special asymptotic series they developed
for the modified Bessel functions which appear in the
standard form of $S_q(x)^{LAT}$. Their result, expressed to ${\cal O}(a)$ is,
\begin{eqnarray} \label{eqn:expandedlattprop}
(4\pi)^2 S_q(x)^{W} \!\!\!&=&\!\!\!
\frac{4m^2}{(x^2)^\frac{1}{2}} \left( 1 +
\frac{ram}{2} \right) K_1\left[
m(x^2)^{\frac{1}{2}}\right]\nonumber\\
\!\!\!&+&\!\!\! 2ram^4 K_2\left[ m{(x^2)}^\frac{1}{2}\right]\nonumber\\
\!\!\!&+&\!\!\! 2ram^4 \frac{\gamma\cdot x}{(x^2)^\frac{1}{2}} K_1\left[
m{(x^2)}^\frac{1}{2}\right]\nonumber\\
\!\!\!&+&\!\!\! 4m^2(1 + ram) \frac{\gamma\cdot x}{x^2}K_2\left[
m{(x^2)}^\frac{1}{2}\right]\nonumber\\
\!\!\!&+&\!\!\!\mathcal{O}(\mathit{a^2}),
\end{eqnarray}
where $K_1$ and $K_2$ are modified Bessel functions of the second kind.

In \cite{abm}, $S_q(x)^{W}$ was used instead of Eq.(\ref{eqn:qprop})
in Eq.(\ref{eq:g2}) thereby obtaining an expression for $G_2(t)$
for Wilson fermions correct to ${\cal O}(a)$. \cite{abm} lists results
for all
of the commonly used mesonic channels. Here we list only the result for
the vector channel:
\begin{eqnarray} \nonumber
G_2^{Vi}(t)^{OPE,W}
 &=& \frac{-1}{2 \pi^2 t^3}[1 + ra (m_1+m_2)] \\ \nonumber
 &+& \frac{3}{8 \pi^2 t} [ m_1^2 + m_2^2 - 2 m_1 m_2 \\
 & & + 2ra(m_1 m_2^2 + m_1^2 m_2) ],
\label{eq:vi_wf}
\end{eqnarray}
where $r$ is the Wilson parameter.
Note that, as expected, $G_2^{OPE,W} = G_2^{OPE,cont} + {\cal O}(a)$.

Table \ref{tb:fits} shows the results for the lattice-QCD Continuum
Model fit, Eq.(\ref{eq:vi_wf}). In this table the $\chi^2$ values
are shown together with the estimates of the $Z$ and $M$ values.
The lattice-QCD Continuum Model fits are very similar to the QCD
Continuum Model fits from Eq.(\ref{eq:vi_hl}), but they
become significantly better (i.e. have a smaller
$\chi^2$) as the quark mass increases.
This is to be expected since the lattice artefacts $\sim ma$.
(Note that in Figure \ref{fig:vi_hl} where the effective mass for the vector
channel at $\kappa_{1,2} = 0.1385$ is plotted, the results from the
fit to Eq.(\ref{eq:vi_wf}) is indistinguishable by eye to that from
Eq.(\ref{eq:vi_hl}).)



\section{Maximum Entropy Method and the Nambu--Jona-Lasinio Model}

In the above two sections, an Ansatz for the spectral function
$\rho(s)$ was derived using an analytic expression for the quark
propagator, $S_q(x)$, valid at small $x$. This Ansatz was then
used as a fitting function for Monte Carlo data for 2-point hadronic
functions, $G_2(t)$. It is interesting to ponder if $\rho(s)$ itself
can be directly extracted from Monte Carlo $G_2(t)$ data.

At first sight this aim appears impossible to achieve due to the
intrinsic ambiguity in the inversion. $G_2(t)$ is known at $N_t \approx
{\cal O}(10)$ values of $t$ whereas, in order to reveal spectral structure,
$\rho(s)$ must be calculated at $N_s \gtap {\cal O}(100)$ $s$ values.
This means that there are an infinite number of solutions for $\rho(s)$
which perfectly fit the data $G_2(t)$.

In \cite{ahn} the Maximum Entropy Method (which is well known
in almost every numerical field outside lattice gauge theory!)
was used to overcome this seemingly insurmountable problem.
Bayes' theorem can be used to express the probability of $\rho(s)$
given the data $G_2(t)$, $P(\rho | G_2)$, as
\begin{equation}
P(\rho | G_2) \propto P(G_2 | \rho) \times P(\rho).
\label{eq:bayes}
\end{equation}
Here, $P(G_2 | \rho)$ can be written as the usual exponential
of the standard $\chi^2$ function of the Maximum Likelyhood Method.
The probability, $P(\rho)$ can be expressed in terms of the
entropy function, $S$. The best solution for $\rho$ is obtained
by ``just'' maximising $P(\rho | G_2)$ in Eq.(\ref{eq:bayes}).

Since the work of \cite{ahn} there have been several applications
of the MEM method to lattice simulations. In this paper, we concentrate
on applying the MEM to the Nambu--Jona-Lasinio model in (2+1)D.
This work is discussed in more detail in \cite{njl}.
For details of applying MEM to lattice simulations see \cite{ahn}.

The continuum (Euclidean) NJL Lagrangian is:
\begin{eqnarray*}
{\cal L}= \bar{\Psi}_i(\partial\hskip -.5em / + m_0 + \sigma +
i \gamma_5 \pi)\Psi_i + \frac{N_f}{2 g^{2}} (\sigma^{2}+ \pi^2).
\end{eqnarray*}
The fields $\Psi_i$ and $\bar{\Psi}_i$ are four-component spinors and
the index $i$ runs over $N_f$ fermion flavours.
The model has an interacting continuum limit and has been used
to model the strong interaction. It exists in two phases
defined by a chiral order parameter.
The lattice simulation used the staggered fermion formulation.

\begin{figure}[htb]
\centering
\includegraphics[scale=0.29]{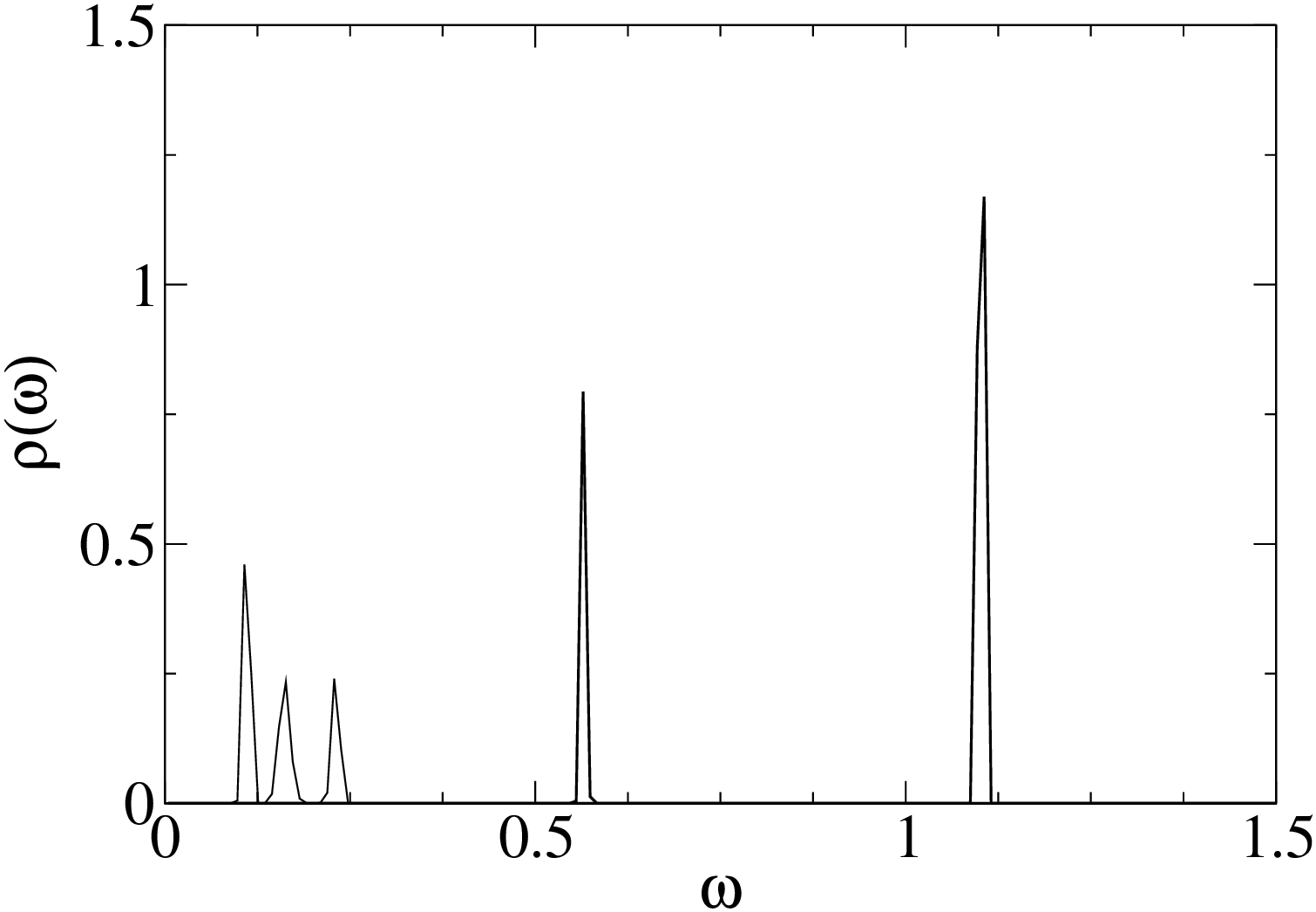}
\caption{\small Broken phase spectral functions for the NJL model.
Peaks from left to right are for the $\pi$ ($m_0=0.005,0.01,0.02$),
f and PS (both $m_0=0.01$).}
\label{fig:mem}
\end{figure}

Figure \ref{fig:mem} shows the spectral function obtained from
MEM for the $\pi$, fermion (f) and massive PS meson in the broken
phase with $N_f=4$ flavours and coupling of $\beta = 1/g^2 = 0.55$.
This plot represents approximately 30,000 configurations of data.
It shows that MEM is easily able to isolate the ground states
in these channels. Furthermore, estimates of the binding energy
of the PS state can be obtained using the MEM spectral functions
which are in agreement with expectations
from large $N_f$ expansions \cite{njl}.

\begin{figure}[htb]
\centering
\includegraphics[scale=0.29]{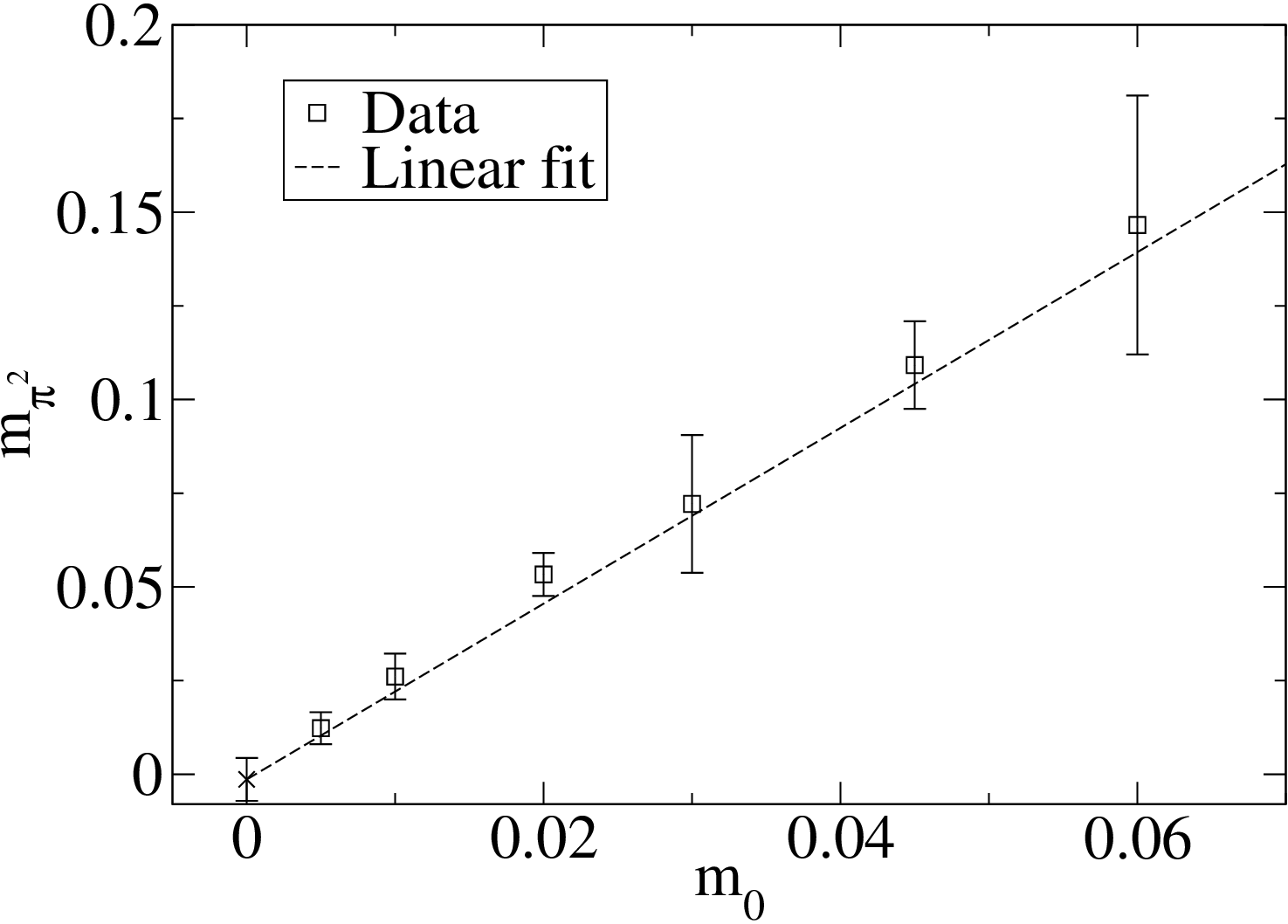}
\caption{\small PCAC scaling relation for $\beta=0.55$ NJL data sets.}
\label{fig:pcac}
\end{figure}

In Figure \ref{fig:pcac}, the $\pi$ mass squared is plotted against
the quark mass showing that the PCAC scaling relationship holds.

Further results for both phases are detailed in \cite{njl}
and future work where MEM is applied to the UKQCD dynamical data set
is detailed in \cite{ukqcd_mem}.



\section{Conclusions}

This paper summarises the results from two interesting
``unconventional'' approaches to the study of 2-point hadronic
correlation functions in lattice Monte Carlo simulations,
$G_2(t)$. The first approach aims at obtaining an accurate analytic
expression for the short distance behaviour of $G_2(t)$ via
the QCD Sum Rules Continuum Model. In particular, the recently
derived expression for the Wilson quark propagator \cite{paladini} was used.
Further work in this area will include the derivation of the analytic
expression for the clover quark propagator and, from that, the
calculation of the analogous expression for $G_2(t)$ at short distances
for this action.

The second approach uses the Maximum Entropy Method to calculate
the spectral function, $\rho(s)$, {\em given} the 2-point function,
$G_2(t)$. This technique is being used by a growing
number of researchers in this context. In this work we used MEM to study the 
Nambu--Jona-Lasinio model and found it to work admirably.



\section{Acknowledgements}

The authors wish to thank Simon Hands, Craig McNeile and Costas Strouthos
for invaluable help
and they acknowledge the UKQCD Collaboration for use of Cray T3E resources.




\end{document}